# Study of $t\bar{t}$ Production in $p\bar{p}$ Collisions Using Total Transverse Energy

F. Abe,[14] H. Akimoto,[32] A. Akopian,[27] M. G. Albrow,[7] S. R. Amendolia,[24] D. Amidei,[17] J. Antos,[29] C. Anway-Wiese,[4] S. Aota,[32] G. Apollinari,[27] T. Asakawa,[32] W. Ashmanskas,[15] M. Atac,[7] P. Auchincloss,[26] F. Azfar,[22] P. Azzi-Bacchetta,[21] N. Bacchetta,[21] W. Badgett,[17] S. Bagdasarov,[27] M. W. Bailey,[19] J. Bao,[35] P. de Barbaro,[26] A. Barbaro-Galtieri,[15] V. E. Barnes,[25] B. A. Barnett,[13] P. Bartalini,[24] G. Bauer,[16] T. Baumann,[9] F. Bedeschi,[24] S. Behrends,[3] S. Belforte,[24] G. Bellettini,[24] J. Bellinger,[34] D. Benjamin,[31] J. Benlloch,[16] J. Bensinger,[3] D. Benton,[22] A. Beretvas,[7] J. P. Berge,[7] S. Bertolucci,[8] A. Bhatti,[27] K. Biery,[12] M. Binkley,[7] D. Bisello,[21] R. E. Blair,[1] C. Blocker,[3] A. Bodek,[26] W. Bokhari,[16] V. Bolognesi,[24] D. Bortoletto,[25] J. Boudreau,[23] L. Breccia,[2] C. Bromberg,[18] E. Buckley-Geer,[7] H. S. Budd,[26] K. Burkett,[17] G. Busetto,[21] A. Byon-Wagner,[7] K. L. Byrum,[1] J. Cammerata,[13] C. Campagnari,[7] M. Campbell,[17] A. Caner,[7] W. Carithers,[15] D. Carlsmith,[34] A. Castro,[21] D. Cauz,[24] Y. Cen,[26] F. Cervelli,[24] H. Y. Chao,[29] J. Chapman,[17] M.-T. Cheng,[29] G. Chiarelli,[24] T. Chikamatsu,[32] C. N. Chiou,[29] L. Christofek,[11] S. Cihangir,[7] A. G. Clark,[24] M. Cobal,[24] M. Contreras,[5] J. Conway,[28] J. Cooper,[7] M. Cordelli,[8] C. Couyoumtzelis,[24] D. Crane,[1] D. Cronin-Hennessy,[6] R. Culbertson,[5] J. D. Cunningham,[3] T. Daniels,[16] F. DeJongh,[7] S. Delchamps,[7] S. Dell'Agnello,[24] M. Dell'Orso,[24] L. Demortier,[27] B. Denby,[24] M. Deninno,[2] P. F. Derwent,[17] T. Devlin,[28] M. Dickson,[26] J. R. Dittmann,[6] S. Donati,[24] J. Done,[30] R. B. Drucker,[15] A. Dunn,[17] N. Eddy,[17] K. Einsweiler,[15] J. E. Elias,[7] R. Ely,[15] E. Engels, Jr.,[23] D. Errede,[11] S. Errede,[11] Q. Fan,[26] I. Fiori,[2] B. Flaugher,[7] G. W. Foster,[7] M. Franklin,[9] M. Frautschi,[19] J. Freeman,[7] J. Friedman,[16] H. Frisch,[5] T. A. Fuess,[1] Y. Fukui,[14] S. Funaki,[32] G. Gagliardi,[24] S. Galeotti,[24] M. Gallinaro,[21] M. Garcia-Sciveres,[15] A. F. Garfinkel,[25] C. Gay,[9] S. Geer,[7] D. W. Gerdes,[17] P. Giannetti,[24] N. Giokaris,[27] P. Giromini,[8] L. Gladney,[22] D. Glenzinski,[13] M. Gold,[19] J. Gonzalez,[22] A. Gordon,[9] A. T. Goshaw,[6] K. Goulianos,[27] H. Grassmann,[7,*] L. Groer,[28] C. Grosso-Pilcher,[5] G. Guillian,[17] R. S. Guo,[29] C. Haber,[15] S. R. Hahn,[7] R. Hamilton,[9] R. Handler,[34] R. M. Hans,[35] K. Hara,[32] A. D. Hardman,[25] B. Harral,[22] R. M. Harris,[7] S. A. Hauger,[6] J. Hauser,[4] C. Hawk,[28] E. Hayashi,[32] J. Heinrich,[22] K. D. Hoffman,[25] M. Hohlmann,[1,5] C. Holck,[22] R. Hollebeek,[22] L. Holloway,[11] A. Hölscher,[12] S. Hong,[17] G. Houk,[22] P. Hu,[23] B. T. Huffman,[23] R. Hughes,[26] J. Huston,[18] J. Huth,[9] J. Hylen,[7] H. Ikeda,[32] M. Incagli,[24] J. Incandela,[7] J. Iwai,[32] Y. Iwata,[10] H. Jensen,[7] U. Joshi,[7] R. W. Kadel,[15] E. Kajfasz,[7,*] T. Kamon,[30] T. Kaneko,[32] K. Karr,[33] H. Kasha,[35] Y. Kato,[20] L. Keeble,[8] K. Kelley,[16] R. D. Kennedy,[28] R. Kephart,[7] P. Kesten,[15] D. Kestenbaum,[9] R. M. Keup,[11] H. Keutelian,[7] F. Keyvan,[4] B. J. Kim,[26] D. H. Kim,[7,*] H. S. Kim,[12] S. B. Kim,[17] S. H. Kim,[32] Y. K. Kim,[15] L. Kirsch,[3] P. Koehn,[26] K. Kondo,[32] J. Konigsberg,[9] S. Kopp,[5] K. Kordas,[12] W. Koska,[7] E. Kovacs,[7,*] W. Kowald,[6] M. Krasberg,[17] J. Kroll,[7] M. Kruse,[25] T. Kuwabara,[32] S. E. Kuhlmann,[1] E. Kuns,[28] A. T. Laasanen,[25] N. Labanca,[24] S. Lammel,[7] J. I. Lamoureux,[3] T. LeCompte,[11] S. Leone,[24] J. D. Lewis,[7] P. Limon,[7] M. Lindgren,[4] T. M. Liss,[11] N. Lockyer,[22] O. Long,[22] C. Loomis,[28] M. Loreti,[21] J. Lu,[30] D. Lucchesi,[24] P. Lukens,[7] S. Lusin,[34] J. Lys,[15] K. Maeshima,[7] A. Maghakian,[27] P. Maksimovic,[16] M. Mangano,[24] J. Mansour,[18] M. Mariotti,[21] J. P. Marriner,[7] A. Martin,[11] J. A. J. Matthews,[19] R. Mattingly,[16] P. McIntyre,[30] P. Melese,[27] A. Menzione,[24] E. Meschi,[24] S. Metzler,[22] C. Miao,[17] G. Michail,[9] S. Mikamo,[14] R. Miller,[18] H. Minato,[32] S. Miscetti,[8] M. Mishina,[14] H. Mitsushio,[32] T. Miyamoto,[32] S. Miyashita,[32] Y. Morita,[14] J. Mueller,[23] A. Mukherjee,[7] T. Muller,[4] P. Murat,[24] H. Nakada,[32] I. Nakano,[32] C. Nelson,[7] D. Neuberger,[4] C. Newman-Holmes,[7] M. Ninomiya,[32] L. Nodulman,[1] S. H. Oh,[6] K. E. Ohl,[35] T. Ohmoto,[10] T. Ohsugi,[10] R. Oishi,[32] M. Okabe,[32] T. Okusawa,[20] R. Oliver,[22] J. Olsen,[34] C. Pagliarone,[2] R. Paoletti,[24] V. Papadimitriou,[31] S. P. Pappas,[35] S. Park,[7] J. Patrick,[7] G. Pauletta,[24] M. Paulini,[15] L. Pescara,[21] M. D. Peters,[15] T. J. Phillips,[6] G. Piacentino,[2] M. Pillai,[26] K. T. Pitts,[7] R. Plunkett,[7] L. Pondrom,[34] J. Proudfoot,[1] F. Ptohos,[9] G. Punzi,[24] K. Ragan,[12] A. Ribon,[21] F. Rimondi,[2] L. Ristori,[24] W. J. Robertson,[6] T. Rodrigo,[7,*] J. Romano,[5] L. Rosenson,[16] R. Roser,[11] W. K. Sakumoto,[26] D. Saltzberg,[5] L. Santi,[24] H. Sato,[32] V. Scarpine,[30] P. Schlabach,[9] E. E. Schmidt,[7] M. P. Schmidt,[35] G. F. Sciacca,[24] A. Scribano,[24] S. Segler,[7] S. Seidel,[19] Y. Seiya,[32] G. Sganos,[12] A. Sgolacchia,[2] M. D. Shapiro,[15] N. M. Shaw,[25] Q. Shen,[25] P. F. Shepard,[23] M. Shimojima,[32] M. Shochet,[5] J. Siegrist,[15] A. Sill,[31] P. Sinervo,[12] P. Singh,[23] J. Skarha,[13] K. Sliwa,[33] D. A. Smith,[24] F. D. Snider,[13] T. Song,[17] J. Spalding,[7] P. Sphicas,[16] M. Spiropulu,[9] L. Spiegel,[7] A. Spies,[13] L. Stanco,[21] J. Steele,[34] A. Stefanini,[24] K. Strahl,[12] J. Strait,[7] R. Stroehmer,[9] D. Stuart,[7] G. Sullivan,[5] A. Soumarokov,[29] K. Sumorok,[16] J. Suzuki,[32] T. Takada,[32] T. Takahashi,[20] T. Takano,[32] K. Takikawa,[32] N. Tamura,[10] F. Tartarelli,[24] W. Taylor,[12] P. K. Teng,[29] Y. Teramoto,[20] S. Tether,[16] D. Theriot,[7] T. L. Thomas,[19] R. Thun,[17] M. Timko,[33] P. Tipton,[26] A. Titov,[27] S. Tkaczyk,[7] D. Toback,[5] K. Tollefson,[26] A. Tollestrup,[7] J. Tonnison,[25] J. F. de Troconiz,[9] S. Truitt,[17] J. Tseng,[13] N. Turini,[24] T. Uchida,[32] N. Uemura,[32] F. Ukegawa,[22] G. Unal,[22] S. C. van den Brink,[23] S. Vejcik, III,[17] G. Velev,[24] R. Vidal,[7] M. Vondracek,[11] D. Vucinic,[16] R. G. Wagner,[1] R. L. Wagner,[7] J. Wahl,[5] R. C. Walker,[26] C. Wang,[6] C. H. Wang,[29] G. Wang,[24] J. Wang,[5] M. J. Wang,[29] Q. F. Wang,[27] A. Warburton,[12] G. Watts,[26] T. Watts,[28] R. Webb,[30] C. Wei,[6] C. Wendt,[34] H. Wenzel,[15] W. C. Wester, III,[7] A. B. Wicklund,[1] E. Wicklund,[7] R. Wilkinson,[22] H. H. Williams,[22] P. Wilson,[5] B. L. Winer,[26] D. Wolinski,[17] J. Wolinski,[30] X. Wu,[24] J. Wyss,[21] A. Yagil,[7] W. Yao,[15] K. Yasuoka,[32] Y. Ye,[12] G. P. Yeh,[7] P. Yeh,[29] M. Yin,[6] J. Yoh,[7] C. Yosef,[18] T. Yoshida,[20] D. Yovanovitch,[7] I. Yu,[35] J. C. Yun,[7] A. Zanetti,[24] F. Zetti,[24] L. Zhang,[34] W. Zhang,[22] and S. Zucchelli[2]

(CDF Collaboration)

[1] Argonne National Laboratory, Argonne, Illinois 60439
[2] Istituto Nazionale di Fisica Nucleare, University of Bologna, I-40126 Bologna, Italy
[3] Brandeis University, Waltham, Massachusetts 02254
[4] University of California at Los Angeles, Los Angeles, California 90024
[5] University of Chicago, Chicago, Illinois 60637
[6] Duke University, Durham, North Carolina 27708
[7] Fermi National Accelerator Laboratory, Batavia, Illinois 60510
[8] Laboratori Nazionali di Frascati, Istituto Nazionale di Fisica Nucleare, I-00044 Frascati, Italy
[9] Harvard University, Cambridge, Massachusetts 02138
[10] Hiroshima University, Higashi-Hiroshima 724, Japan
[11] University of Illinois, Urbana, Illinois 61801
[12] Institute of Particle Physics, McGill University, Montreal, Canada H3A 2T8, and University of Toronto, Toronto, Canada M5S 1A7
[13] The Johns Hopkins University, Baltimore, Maryland 21218
[14] National Laboratory for High Energy Physics (KEK), Tsukuba, Ibaraki 305, Japan
[15] Lawrence Berkeley Laboratory, Berkeley, California 94720
[16] Massachusetts Institute of Technology, Cambridge, Massachusetts 02139
[17] University of Michigan, Ann Arbor, Michigan 48109
[18] Michigan State University, East Lansing, Michigan 48824
[19] University of New Mexico, Albuquerque, New Mexico 87131
[20] Osaka City University, Osaka 588, Japan
[21] Universita di Padova, Istituto Nazionale di Fisica Nucleare, Sezione di Padova, I-35131 Padova, Italy
[22] University of Pennsylvania, Philadelphia, Pennsylvania 19104
[23] University of Pittsburgh, Pittsburgh, Pennsylvania 15260
[24] Istituto Nazionale di Fisica Nucleare, University and Scuola Normale Superiore of Pisa, I-56100 Pisa, Italy
[25] Purdue University, West Lafayette, Indiana 47907
[26] University of Rochester, Rochester, New York 14627
[27] Rockefeller University, New York, New York 10021
[28] Rutgers University, Piscataway, New Jersey 08854
[29] Academia Sinica, Taipei, Taiwan 11529, Republic of China
[30] Texas A&M University, College Station, Texas 77843
[31] Texas Tech University, Lubbock, Texas 79409
[32] University of Tsukuba, Tsukuba, Ibaraki 305, Japan
[33] Tufts University, Medford, Massachusetts 02155
[34] University of Wisconsin, Madison, Wisconsin 53706
[35] Yale University, New Haven, Connecticut 06511

We analyze a sample of $W+$ jet events collected with the Collider Detector at Fermilab (CDF) in $p\bar{p}$ collisions at $\sqrt{s} = 1.8$ TeV to study $t\bar{t}$ production. We employ a simple kinematical variable $\mathcal{H}$, defined as the scalar sum of the transverse energies of the lepton, neutrino and jets. For events with a $W$ boson and four or more jets, the shape of the $\mathcal{H}$ distribution deviates by 3.8 standard deviations from that expected from known backgrounds to $t\bar{t}$ production. However this distribution agrees well with a linear combination of background and $t\bar{t}$ events, the agreement being best for a top mass of 180 GeV/c$^2$.

PACS numbers: 14.65.Ha, 13.85.Ni, 13.85.Qk

The existence of the top quark has recently been established by the CDF [1,2] and D0 [3] collaborations. In the CDF analyses [1,2], $b$ quark tagging was used to select $t\bar{t}$ candidates in a sample of $W+\geq 3$-jet events, where the $W$ decays into $e\nu$ or $\mu\nu$. A $W+\geq 4$-jet subsample was then used to reconstruct the top mass under the hypothesis that the top quark decays into a $W$ boson and a $b$ quark. The mass peak in the $b$-tagged $W+$ jet data gives evidence for the top quark using a kinematic variable. In addition a study [4] of the transverse energy ($E_T$) [5] spectrum of the second and third highest $E_T$ jets was able to identify a contribution of $t\bar{t}$ production in our data.

For this analysis we use the CDF $W+\geq 4$-jet subsample and study the variable $\mathcal{H}$, which is defined as the scalar sum of the lepton $E_T$, the neutrino $E_T$ as measured by the missing-$E_T$ ($\not\!\!E_T$) in the event, and the $E_T$ of each jet [6,7]. The variable $\mathcal{H}$ is strongly correlated with the center of mass energy ($\sqrt{\hat{s}}$) of the parton-parton hard scattering process. In $t\bar{t}$ events, it is also correlated to the transverse mass [8] of the $t\bar{t}$ system. For a top mass larger than about 140 GeV/c$^2$, $\mathcal{H}$ has discriminating power between $t\bar{t}$ and $W+$ multijet background events, because the $W$ and $b$ quark resulting from the top quark decay have higher transverse momenta ($P_T$) than radiated gluons in background processes. For this study no $b$-tagging or top event reconstruction is required. Therefore, this analysis has larger acceptance for a $t\bar{t}$ signal and has different systematic uncertainties. It is not affected by ambiguities in jet-parton assignments, as is the case for the event-fitting algorithm used to determine the top mass in Ref. [2] and [3]. In addition, it is sensitive to unconventional top decay modes which do not involve $b$ quarks, such as $t \rightarrow Ws$ or $t \rightarrow Wd$. In these respects the study of $\mathcal{H}$ supplements Ref. [2] by providing additional information on the top quark mass and the $t\bar{t}$ production rate.

We report here on a study of the $W+\geq$ 4-jet data sample used in Ref. [2] and based on 67 pb$^{-1}$ of integrated luminosity. We search for $t\bar{t}$ events where one $W$ decays leptonically into a $\ell\nu$ pair ($\ell = e, \mu$), and the other $W$ decays into quarks. The CDF detector and the requirements on the lepton and $\not\!\!E_T$ are described in Refs. [1,2] and result in approximately 71,500 events. We then require events to have 3 jets that pass "high-threshold" cuts, with the observed calorimeter $E_T \geq$ 15 GeV and $|\eta_{jet}| \leq 2.0$, and at least one additional jet that passes a "low-threshold" cut of observed $E_T \geq 8$ GeV and $|\eta_{jet}| \leq 2.4$. These jet cuts provide the 99-event signal sample used for this paper and for the mass analysis in Ref. [2], out of which 88 events had a good reconstructed fit to the $t\bar{t}$ hypothesis. When calculating $\mathcal{H}$, the $\not\!\!E_T$ and the $E_T$ of the above 4 jets are corrected for detector response. For 35% of the events in the $W+\geq$ 4-jet sample, there are extra jets that pass the low-threshold cuts and the corrected $E_T$ of these extra jets are added to the $\mathcal{H}$ variable.

The dominant background to $t\bar{t}$ production in the $W$+jets mode is direct production of a $W$ which recoils against energetic light quarks and gluons. Other backgrounds involving real $W$ bosons are: $WW$, $WZ$+jets where a $W$ decays into $\ell\nu$; and $W$+jet events where $W \rightarrow \tau\nu$ and $\tau \rightarrow \ell\nu_\ell\nu_\tau$. The non-$W$ backgrounds are: QCD multijet production where one jet fakes an electron or a muon (QCD fakes); $b\bar{b}$ + multijet production where one of the $b$ quarks decays semileptonically; $ZZ$+jets with one $Z$ decaying leptonically but with only one lepton found; $Z \rightarrow \tau\tau$ followed by $\tau \rightarrow \ell\nu_\ell\nu_\tau$; and Drell-Yan ($\gamma^*, Z^0$) production of lepton pairs along with extra QCD jets.

We model the shape of $\mathcal{H}$ for the $W+$ jets background using the VECBOS [9] Monte Carlo program. From the data and various Monte Carlo estimates, we obtain the $\mathcal{H}$ spectrum for all other background processes and find it is well matched to the VECBOS $\mathcal{H}$ spectrum. The main systematic uncertainty in the VECBOS calculation is the $Q^2$ scale, which determines the strong coupling constant and therefore the production rate and the shape of the kinematic distributions. We use two choices of $Q^2$ scale, the square of the average $P_T$ of the jets ($\langle P_T \rangle^2$) and the square of the $W$ boson mass ($M_W^2$). The $t\bar{t}$ Monte Carlo samples are generated with the HERWIG program [10]. All the Monte Carlo samples are processed by a detector simulation program and reconstructed in the same way as the data. We emphasize that this analysis is based on comparing the shape of the measured $\mathcal{H}$ distribution, not the rate, with the signal and background predictions.

TABLE I. Definition of the $W+$ jets control and signal samples. The third, fourth and fifth columns list the criteria placed on the jets in each event.

| Sample | Threshold | $N_{jets}$ | $E_T$ cut | $\|\eta_{jet}\|$ cut | Events |
|---|---|---|---|---|---|
| Control | low | $= 3$ | $\geq 8$ | $\leq 2.4$ | 814 |
| | | Veto jet 4 | $\geq 8$ | $\leq 2.4$ | |
| Control | high | $= 3$ | $\geq 15$ | $\leq 2.0$ | 104 |
| | | Veto jet 4 | $\geq 8$ | $\leq 2.4$ | |
| Signal | low | $\geq 4$ | $\geq 8$ | $\leq 2.4$ | 267 |
| Signal | high | $\geq 3$ | $\geq 15$ | $\leq 2.0$ | 99 |
| | | $\geq 1$ | $\geq 8$ | $\leq 2.4$ | |

To test how well the VECBOS calculation models the background, we define two top-depleted $W+$3-jet control samples, and a low-threshold $W+\geq$ 4-jet signal sample in addition to the high-threshold signal sample defined above. These 4 samples are summarized in Table I. The $t\bar{t}$ contamination in the control samples is expected to be about 1% (10%) for the low- (high-) threshold data [1], assuming a top mass of 175 GeV/c$^2$. These two control samples provide a means of verifying the $W+$ 3-jet background calculation. The low-threshold signal sample with $\mathcal{H} < 200$ GeV is depleted in top events and allows us to test the $W+\geq$ 4-jet background calculation.

We first compare the data and the VECBOS prediction for our control samples. The $\mathcal{H}$ plots for both control samples are shown in Fig. 1, along with the VECBOS calculations for $Q^2 = \langle P_T \rangle^2$ mixed with the expected top component (1% top in Fig. 1a and 10% top in Fig. 1b). The Monte Carlo distributions describe the data well. For $Q^2 = M_W^2$, we find that the low threshold VECBOS distribution is slightly shifted to larger values of $\mathcal{H}$. However, the size of this shift is similar to the uncertainty in $\mathcal{H}$ from the 10% systematic error in the jet energy scale [1]. We therefore cannot reject the $Q^2 = M_W^2$ scale on the basis of this comparison. The high-threshold data agree equally well with the VECBOS predictions for $Q^2 = \langle P_T \rangle^2$ and $Q^2 = M_W^2$. We conclude that the VECBOS calculation, plus the small $t\bar{t}$ contamination,

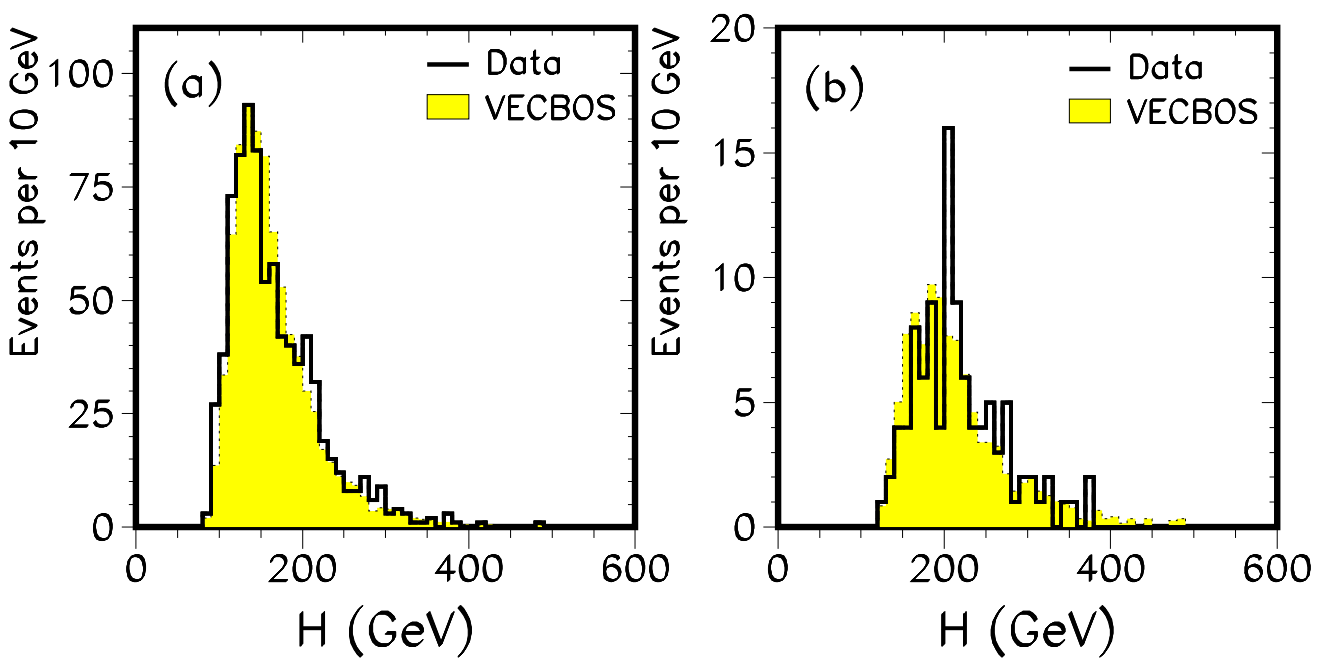

FIG. 1. Comparison of $\mathcal{H}$ distributions for the control data (solid line) and the VECBOS ($Q^2 = \langle P_T \rangle^2$) prediction (shaded) for (a) $W+$ 3-jet events passing the low-threshold cuts, and (b) $W+$3-jet events passing the high-threshold cuts. The VECBOS prediction plus 1% top for (a) and 10% top for (b), has been normalized to the data.

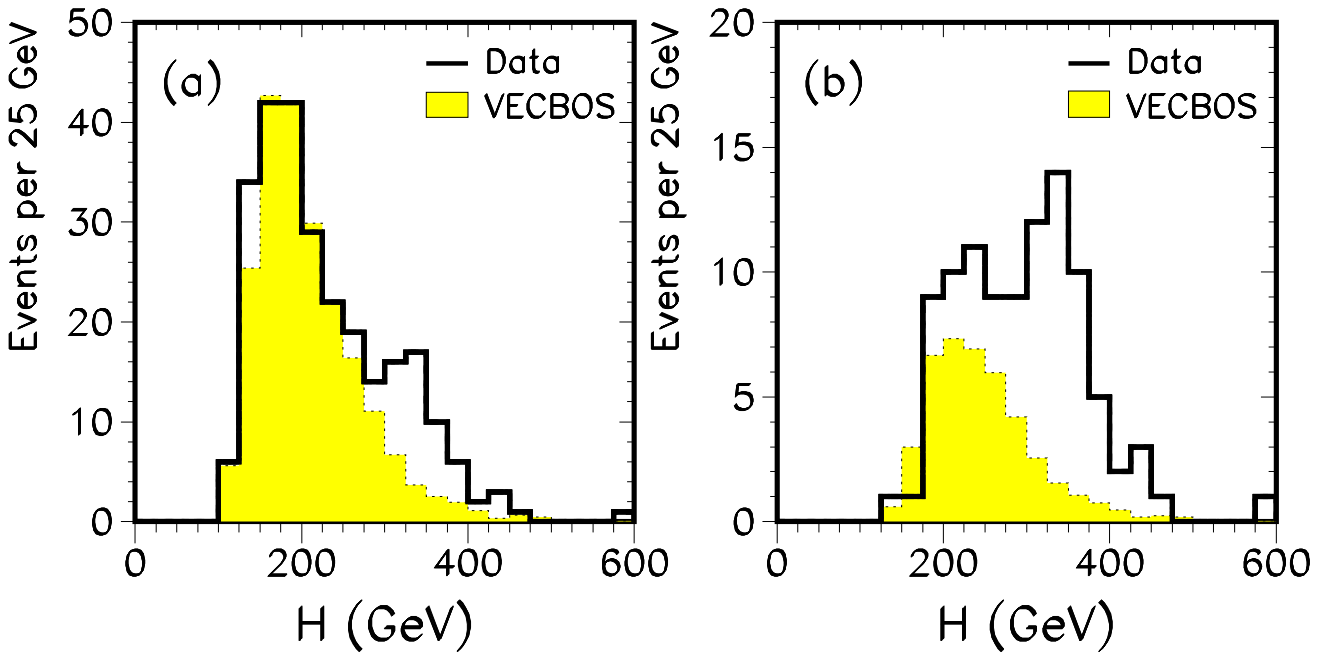

FIG. 2. Comparison of $\mathcal{H}$ distributions for the signal data (solid line) and the VECBOS ($Q^2 = \langle P_T \rangle^2$) Monte Carlo prediction (shaded) for (a) $W+ \geq$ 4-jet events passing the low-threshold cuts, and (b) $W+ \geq$ 4-jet events passing the high-threshold cuts. The Monte Carlo prediction is normalized to the expected number of background events obtained by the fitting procedure described in the text.

describes both the low- and high-threshold control data well within our systematic uncertainties.

We next compare the data and the VECBOS prediction for our signal samples. The $\mathcal{H}$ plot for $W+ \geq$ 4-jet events passing the low-threshold cuts is shown in Fig. 2a, together with the VECBOS prediction for $Q^2 = \langle P_T \rangle^2$. The VECBOS distribution is normalized to the expected number of events obtained by the fit described in the next paragraph. The data and VECBOS prediction agree well in both the peak position and the shape for $\mathcal{H}$ below 200 GeV, suggesting that the low-$\mathcal{H}$ events are predominantly background. However, on the high-$\mathcal{H}$ side above 200 GeV, a shoulder is seen in the data above the VECBOS curve. The histogram of $\mathcal{H}$ for the 99 $W+$ jet events passing the high-threshold cuts is shown in Fig. 2b. This sample consists of less than half the events in the low-threshold sample and is expected to include only 5% less top. The corresponding VECBOS distribution with $Q^2 = \langle P_T \rangle^2$ is also shown, normalized in the same way as for the low-threshold sample. The data is significantly broader than the VECBOS prediction. Performing an unbinned Kolmogorov-Smirnov test, we find that the probability for the $Q^2 = \langle P_T \rangle^2$ VECBOS prediction to fluctuate to the observed data is $1.8 \times 10^{-9}$ ($6\sigma$ for a Gaussian distribution). When the VECBOS calculation with $Q^2 = M_W^2$ is used, this probability is reduced to $3.6 \times 10^{-6}$ ($4.6\sigma$). In the conservative case where $Q^2 = M_W^2$ is used and the $E_T$ of each jet is increased by 10% in the VECBOS program, the probability that the background is consistent with the $W+ \geq$ 4-jet data is $1.6 \times 10^{-4}$ ($3.8\sigma$).

To study the high-$\mathcal{H}$ events in the $W+ \geq$ 4-jet sample, we perform binned maximum likelihood fits [11] of the data to a linear combination of the $\mathcal{H}$ distributions predicted by the VECBOS and the $t\bar{t}$ HERWIG Monte Carlo calculations. Figure 3 shows the negative log-likelihoods for the two-component fits as a function of the top quark mass ($M_{TOP}$). After fitting the data points in this plot to a cubic polynomial, we find the top quark mass to be $180 \pm 12(\text{stat.})^{+19}_{-15}(\text{syst.})$ GeV/c$^2$. The systematic error reflects our uncertainties in the jet energy scale, the $Q^2$ scale in VECBOS, and the level of initial state radiation predicted by the HERWIG $t\bar{t}$ calculation. This top quark mass value is in excellent agreement with that found by the mass analysis of the $b$-tagged $W+ \geq$ 4-jet events [2]. The two estimates are correlated since the 19 events in Ref. [2] are a subset of the 99 events used here. Because of this, we only quote the $M_{TOP}$ estimate from this analysis to demonstrate its consistency with our earlier result of $176 \pm 8 \pm 10$ GeV/c$^2$, which remains our best measurement for $M_{TOP}$. For a top mass of 180 GeV/c$^2$, and setting $Q^2 = \langle P_T \rangle^2$ in the VECBOS calculation, the fit yields $56 \pm 10(\text{stat.}) \pm 5(\text{sys.})$ $t\bar{t}$ events in the high-threshold signal sample of 99 events. The corresponding number with $Q^2 = M_W^2$ is $45 \pm 11(\text{stat.}) \pm 5(\text{sys.})$. This can be compared with $34 \pm 10(\text{stat.}) \pm 5(\text{sys.})$ events as extrapolated from the number of $b$-tagged $W+ \geq$ 3-jet events reported in [2]. These three estimates of the $t\bar{t}$ production rate are consistent within errors. Table II summarizes the results of our analysis.

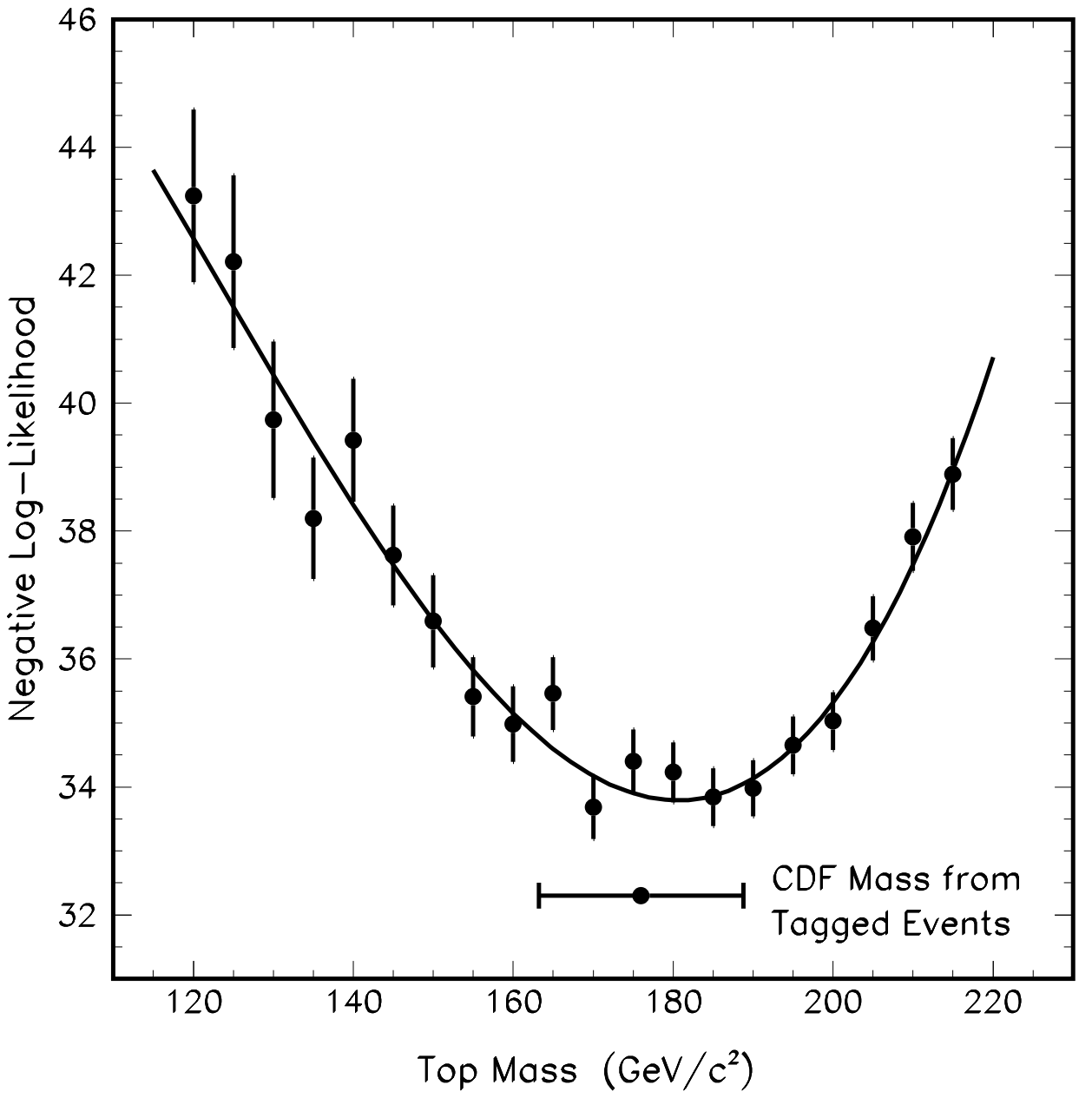

FIG. 3. The least-squares fit of a cubic polynomial to the negative log-likelihood values from the two-component fits (see text), versus the top quark mass. The VECBOS $Q^2$ scale is set to $\langle P_T \rangle^2$. The error bars reflect the statistical uncertainties of the fit due to finite Monte Carlo event samples. Also shown is the CDF mass result and error of Ref. [2].

We also perform a binned maximum likelihood fit to the low-threshold signal sample. The extracted top mass and the number of $t\bar{t}$ events agree well with the results of the high-threshold fit, within statistical uncertainties.

Figure 4 shows a single fit to the high-threshold signal sample assuming a top quark mass of 180 GeV/c$^2$. The $Q^2$ scale was set to $\langle P_T \rangle^2$ in the VECBOS prediction.

TABLE II. The fit results from the high-threshold signal sample for two choices of the $Q^2$ scale in VECBOS. The first error is statistical and the second error is systematic. The $K.S.$ column is a Kolmogorov-Smirnov test confidence level.

| $Q^2$ | $M_{TOP}$ (GeV/c$^2$) | $N_{t\bar{t}}$ | $K.S.$ (%) |
|---|---|---|---|
| $\langle P_T\rangle^2$ | $180 \pm 12^{+19}_{-15}$ | $56 \pm 10 \pm 5$ | 56.5 |
| $M_W^2$ | $184 \pm 15^{+19}_{-15}$ | $45 \pm 11 \pm 5$ | 63.2 |

Good agreement between the best fit and the data is observed, as indicated by the Kolmogorov-Smirnov confidence levels shown in Table II. Also shown are the 23 events tagged by the Silicon Vertex Detector (SVX) and the Soft Lepton Tagging (SLT) algorithms, out of which 19 events had a good reconstructed fit to the $t\bar{t}$ hypothesis in Ref. [2]. Most of the $b$-tags lie in the region $\mathcal{H} \geq 270$ GeV, which as the two-component fit shows, is dominated by $t\bar{t}$ events. A study of background and simulated $t\bar{t}$ events shows that the dependence of the tagging efficiency on $\mathcal{H}$ is negligible. The clustering of $b$-tagged events at large $\mathcal{H}$ provides additional evidence that the excess of events above the $W+$ jet background curve results from $t\bar{t}$ production.

In summary, we compare the total transverse energy ($\mathcal{H}$) distribution of $W+\geq$ 4-jet data with that expected from all known backgrounds and establish that they do not result from the same physical processes. We interpret the excess of high-$\mathcal{H}$ events observed in the data as the result of $t\bar{t}$ production. The best fit to the data is a linear combination of $W+$ jet and $t\bar{t}$ Monte Carlo calculations for a top mass of 180 GeV/c$^2$. We also find a large fraction of events with at least one $b$-tag in the $t\bar{t}$-enhanced high-$\mathcal{H}$ region, consistent with the standard model prediction that one of the top quark decay products is the $b$ quark. We demonstrate that a purely kinematic variable can be used to measure the top mass and the $t\bar{t}$ production rate, and that $\mathcal{H}$ can be used in future analyses to discriminate against backgrounds to the top signal.

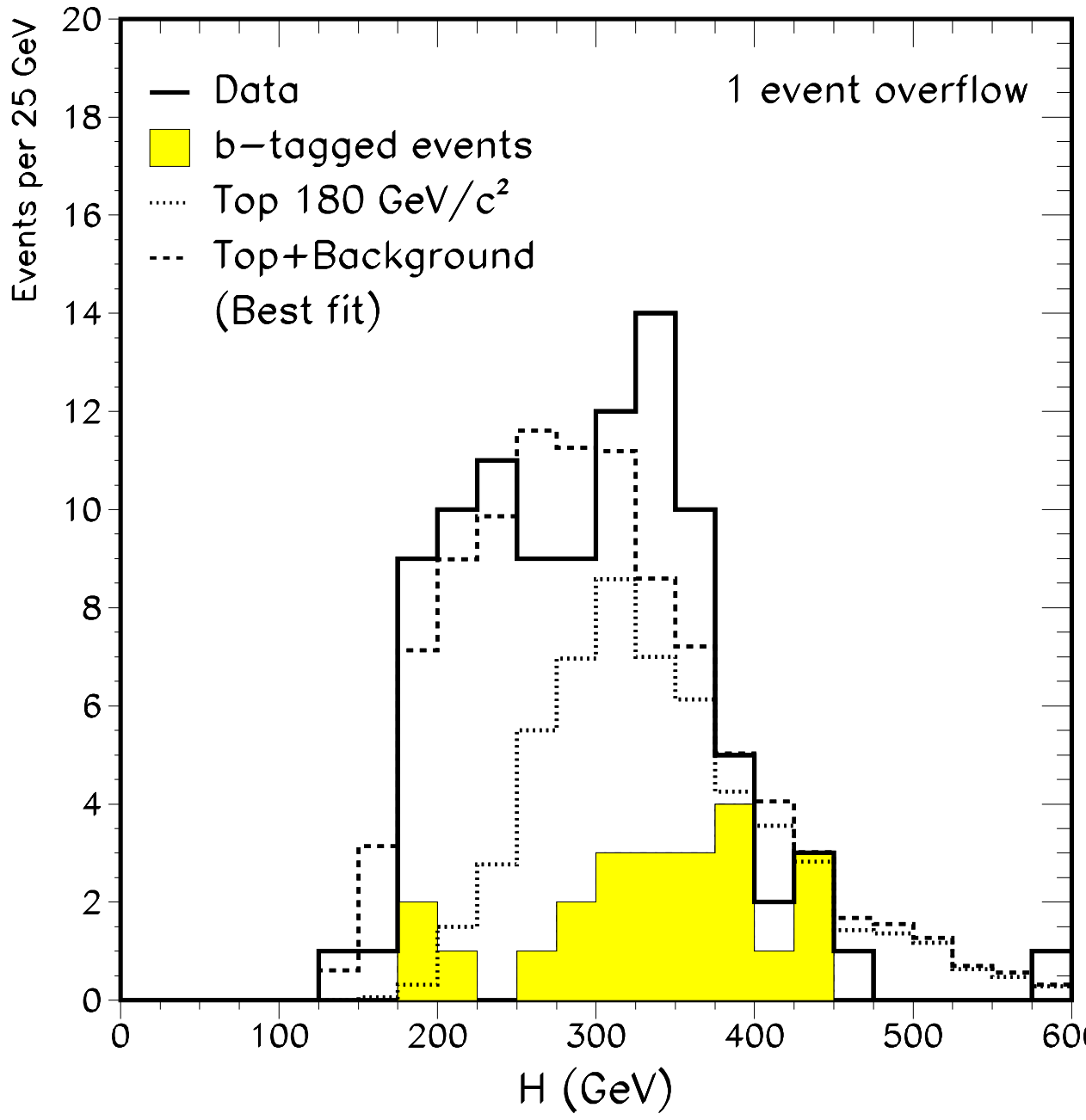

FIG. 4. The binned maximum likelihood fit of the high-threshold signal sample (solid line) to a linear combination of the VECBOS $W+\geq$ 4-jet and HERWIG $t\bar{t}$ predictions (dashed line), for $M_{TOP}$ = 180 GeV/c$^2$. The dotted histogram is the $t\bar{t}$ component of the fit. The $\mathcal{H}$ distribution for the SVX and SLT tagged events is also shown (shaded).

We thank the Fermilab staff and the technical staffs of the participating institutions for their vital contributions. This work is supported by the U.S. Department of Energy and the National Science Foundation, the Natural Sciences and Engineering Research Council of Canada, the Istituto Nazionale di Fisica Nucleare of Italy, the Ministry of Education, Science and Culture of Japan, the National Science Council of the Republic of China, and the A.P. Sloan Foundation.

* Visitor.

[1] F. Abe *et al.*, Phys. Rev. **D50**, 2966 (1994); F. Abe *et al.*, Phys. Rev. Lett. **73**, 225 (1994).

[2] F. Abe *et al.*, Phys. Rev. Lett. **74**, 2626 (1995).

[3] S. Abachi *et al.*, Phys. Rev. Lett. **74**, 2632 (1995).

[4] F. Abe *et al.*, Phys. Rev. **D51**, 4623 (1995); F. Abe *et al.*, *"Identification of Top Quarks at CDF using Kinematic Variables"*, (Submitted to Phys. Rev. Lett.).

[5] In the CDF coordinate system, $\theta$ is the polar angle with respect to the proton beam direction, and $\phi$ is the azimuthal angle. The pseudorapidity, $\eta$, is defined as $-\ln\tan(\frac{\theta}{2})$. The transverse momentum of a particle is $P_T = P\sin\theta$ where $P$ is the particle momentum. If the magnitude of this vector is obtained using the calorimeter energy rather than the spectrometer momentum, it becomes the transverse energy ($E_T$).

[6] The missing-$E_T$ is defined as the negative of the vector sum of all the transverse energies in the event. In this analysis, jets are reconstructed by a cone-clustering algorithm in $\eta - \phi$ space using a cone radius of 0.4.

[7] We note that variants of the $\mathcal{H}$ variable are proposed by J.M. Benlloch, K. Sumorok, W.T. Giele, Nucl. Phys. **B425**, 3 (1994) and employed by D0 in Ref. [3]. However we include the missing-$E_T$ and the muon $P_T$ (where appropriate) in our definition of $\mathcal{H}$. We also choose to use $\mathcal{H}$ as a continuous variable to estimate the top quark mass and production rate, instead of selecting on $\mathcal{H}$ to reduce the background to $t\bar{t}$ production.

[8] The transverse mass ($M_T$) of 6 particles is defined by $M_T^2 = [\sum_1^6 E_{Ti}]^2 - [\sum_1^6 \vec{P}_{Ti}]^2$

[9] F.A. Berends, W.T. Giele, H.Kuijf, and B. Tausk, Nucl. Phys. **B357**, 32 (1991).

[10] G. Marchesini and B.R. Webber, Nucl. Phys. **B310**, 461 (1988): G. Marchesini *et al.*, Comput. Phys. Commun. **67**, 465 (1992).

[11] R. Barlow and C. Beeston, Comput. Phys. Commun. **77**, 219 (1993).